\documentstyle[11pt,newpasp,twoside,epsf]{article}
\def\edcomment#1{\iffalse\marginpar{\raggedright\sl#1\/}\else\relax\fi}
\reversemarginpar

\begin{document}
\title{Mapping the dark matter using weak lensing}

 \author{Henk Hoekstra}
\affil{Canadian Institute for Theoretical Astrophysics, University of
Toronto, 60 St. George Street, M5S 3H8, Toronto, Canada}
\affil{Department of Astronomy and Astrophysics, University of
Toronto, 60 St. George Street, M5S 3H8, Toronto, Canada}

\begin{abstract}

Weak gravitational lensing of distant galaxies by foreground
structures has proven to be a powerful tool to study the mass
distribution in the universe. The advent of panoramic cameras on 4m
class telescope has led to a first generation of surveys that already
compete with large redshift surveys in terms of the accuracy with
which cosmological parameters can be determined. The next surveys,
which already have started taking data, will provide another major
step forward. At the current level, systematics appear under control,
and it is expected that weak lensing will develop into a key tool in the
era of precision cosmology, provided we improve our knowledge of the
non-linear matter power spectrum and the source redshift distribution.

In this review we will briefly describe the principles of weak lensing
and discuss the results of recent cosmic shear surveys. We show how
the combination of weak lensing and cosmic microwave background
measurements can provide tight constraints on cosmological parameters.
We also demonstrate the usefulness of weak lensing in studies of the
relation between the galaxy distribution and the underlying dark
matter distribution (``galaxy biasing''), which can provide important
constraints on models of galaxy formation. Finally, we discuss new and
upcoming large cosmic shear surveys.

\end{abstract}

\section{Introduction}

The differential deflection of light rays by intervening matter
provides us with a unique way to study the projected mass distribution
along the line of sight, without having to rely on assumptions about
the dynamical state or nature of the deflecting matter. In particular,
the small alignments induced in the shapes of distant galaxies, called
``weak gravitational lensing'' has shown to be a valuable tool in
observational cosmology. It allows us to study the clustering
properties of the dark matter directly (whereas many other techniques
require visible tracers), which allows for a straightforward comparison
with theoretical models of structure formation.

The first succesful applications of weak lensing focussed on massive
objects, such as clusters of galaxies. Recently, with advent of
panoramic cameras, it has become possible to survey large (random)
areas of the sky with the purpose of studying the lensing signal
caused by large scale structure. The first detections of this ``cosmic
shear'' signal were reported in the spring of 2000 (Bacon et al. 2000;
Kaiser et al. 2000; van Waerbeke et al. 2000; Wittman et al. 2000).
Since these early detections, many new results have been published
using a range of telescopes and filters, with varying depth (Brown et
al.  2003; Bacon et al. 2003; Hamana et al. 2003; Hoekstra et
al. 2002a, 2002b; Jarvis et al. 2002; Maoli et al. 2001; Refregier et
al. 2002; Rhodes et al. 2001; van Waerbeke et al. 2001, 2002).

We start by reviewing the quantitities measured in cosmic shear
studies. For a detailed discussion of this subject we refer the reader
of an excellent review by Bartelmann \& Schneider (2001). We proceed
by presenting the results of recent cosmic shear surveys and their
constraints on cosmological parameters. The relation between the
galaxy distribution and the underlying dark matter distribution is
discussed in \S4.  The prospects of this rapidly evolving field are
discussed in \S5.

\section{Lensing by large scale structure}

\subsection{What does the signal mean?}

As photons travel through the universe, they are deflected by mass
fluctuations along the line of sight. As a result the observed images
of distant galaxies are slightly distorted, which gives rise to
coherent galaxy alignments. A key assumption made in weak lensing
studies is that the orientations of the galaxies are random.  This
assumption is expected to break down as models of galaxy formation
suggest that galaxies will align with the large scale gravitational
shear field. However, the amplitude is unknown, and appears not very
important for the current generation of cosmic shear
studies. Upcoming, larger surveys will supress effect of these
intrinsic alignments using photometric redshifts for the source
galaxies.

In addition to a change in shape, the sizes of the images are changed
(magnification). Future surveys might attempt to measure these changes
in size, but currently all cosmic shear results are based on the
measurements of the correlations in the galaxy shapes. In this review
we concentrate on this ``shear'' method.

A key ingredient in the theory of structure formation is the matter
power spectrum. In principle, a succesful measurement of the power
spectrum allows one to place constraints on the nature of the dark
matter particles. The advantage of weak lensing over other methods is
that the observed lensing signal can be related to the matter power
spectrum directly. Most of the recent cosmic shear results are based
on two-point statistics (i.e., matter power spectrum), because it is
the easiest quantity to measure, and for that reason these
measurements will be the focus of this review. We note, whowever, that
the first results based on three-point statistics have been reported
recently (Bernardeau et al. 2002 ; Pen et al. 2003). The latter will
be of great interest with upcoming larger surveys because the
combination of two and three-point statistics allows for an accurate
measure of the matter density $\Omega_m$.

The first step in the lensing analysis is to measure the shapes of the
galaxies. Having done so, one can proceed several ways. For instance,
one can tile the observed fields with apertures and compute the excess
variance (compared to a random field) in the galaxy shapes caused by
weak lensing (top-hat variance), or use a compensated radial weight
function (aperture mass). However, in practice, the weak lensing
surveys have complicated geometries because of masking of regions
around bright stars, etc. As a result the tiling described above is
not practical. Instead, it is better to use the observed ellipticity
correlation functions

\begin{equation}
\xi_{\pm}(\theta)=\langle \gamma_t({\bf x}_i)\gamma_t({\bf x}_j)\rangle  \pm 
\langle \gamma_r({\bf x}_i)\gamma_r({\bf x}_j)\rangle,
\end{equation}

\noindent where $\theta=|{\bf x}_i-{\bf x}_j|$, and $\gamma_t$ and
$\gamma_r$ are the tangential and $45\deg$ rotated shear in the frame
defined by the line connecting the pair of galaxies. This has the
advantage that it uses all information contained in the data (albeit
not necessarily optimal) and does not depend on the survey geometry.
The other two-point statistics can be computed by integrating the
correlation functions with appropriate window functions. Furthermore,
this procedure allows the separation of the signal into two components
(e.g., Crittenden et al. 2002): an ``E''-mode, which is curl-free, and
a ``B''-mode, which is sensitive to the curl of the shear
field. Gravitational lensing arises from a gravitational field, and
hence it is expected to produce a curl-free shear field. Hence, the
``B''-mode can be used as a measure of the residual systematics
(including intrinsic galaxy alignments).

The decompositions of the shear correlation function and the top-hat
variance into ``E'' and ``B''-modes are defined up to a constant. On the
other hand, the decomposition is naturally carried out using the
aperture mass statistic $M_{\rm ap}$, which is defined as

\begin{equation}
M_{\rm ap}(\theta)=\int d^2\phi U(\phi)\kappa(\phi),
\end{equation}

\noindent where $U(\phi)$ is a compensated filter. A detailed
discussion of the use of the aperture mass in cosmic shear studies can
be found in Schneider et al. (1998). In particular, we adopt the
filter function suggested by Schneider et al. (1998) in this review.
With this choice of filter function, the observed variance of the
aperture mass $\langle M_{\rm ap}^2\rangle$ is related to the power
spectrum through

\begin{equation}
\langle M_{\rm ap}^2\rangle(\theta)=2\pi\int_0^\infty dl~l
P_\kappa(l)\left[\frac{12}{\pi (l \theta)^2} J_4(l \theta)\right]^2,
\end{equation}

\noindent where $J_4$ is the fourth-order Bessel function of the first
kind. Ideally one would like to measure the 3d power spectrum, but
gravitational lensing is sensitive to all matter along the line of
sight, and as a result all we can measure is $P_\kappa(l)$, the
convergence power spectrum, defined as

\begin{equation} 
P_\kappa(l)=\frac{9 H_0^4 \Omega_m^2}{4 c^4}
\int\limits_0^{w_H}dw \left(\frac{\bar W(w)}{a(w)}\right)^2
P_\delta\left(\frac{l}{f_K(w)};w\right)~\label{pkappa},
\end{equation}

\noindent where $w$ is the radial (comoving) coordinate, $w_H$
corresponds to the horizon, $a(w)$ the cosmic scale factor, and
$f_K(w)$ the comoving angular diameter distance and $P_\delta$ is
the 3d matter power spectrum.

It is clear from Eq.~\ref{pkappa} that the weak lensing signal
measures the projected power spectrum, weighted by a function of the
redshift distribution of the source galaxies: $\bar W(w)$ is the
source-averaged ratio of angular diameter distances $D_{ls}/D_{s}$ for
a redshift distribution of sources $p_b(w)$:

\begin{equation}
\bar W(w)=\int_w^{w_H} dw' p_b(w')\frac{f_K(w'-w)}{f_K(w')}.
\end{equation}

Hence, it is important to know the redshift distribution of the
sources, in order to relate the observed lensing signal to
$P_\kappa(l)$. Furthermore, in order to infer the 3d power spectrum it
is necessary to deconvolve Eq.~\ref{pkappa}, which is the subject of
various studies. Another complication is the fact that it is necessary
to use the non-linear power spectrum, as has been shown by Jain \&
Seljak (1997). This power spectrum can be computed from the linear
power spectrum following the prescriptions from Peacock \& Dodds
(1996) or Smith et al. (2003). However, the accuracy with which the
non-linear power spectrum is currently known, is already comparable to
the error bars of the largest cosmic shear surveys. Hence we need
to improve our estimates of $P(k)$ on small scales, as well as our
knowledge of the source redshift distributions.

\begin{figure}[!t]
\begin{center}
\hbox{
\epsfxsize=0.48\hsize
\epsfbox{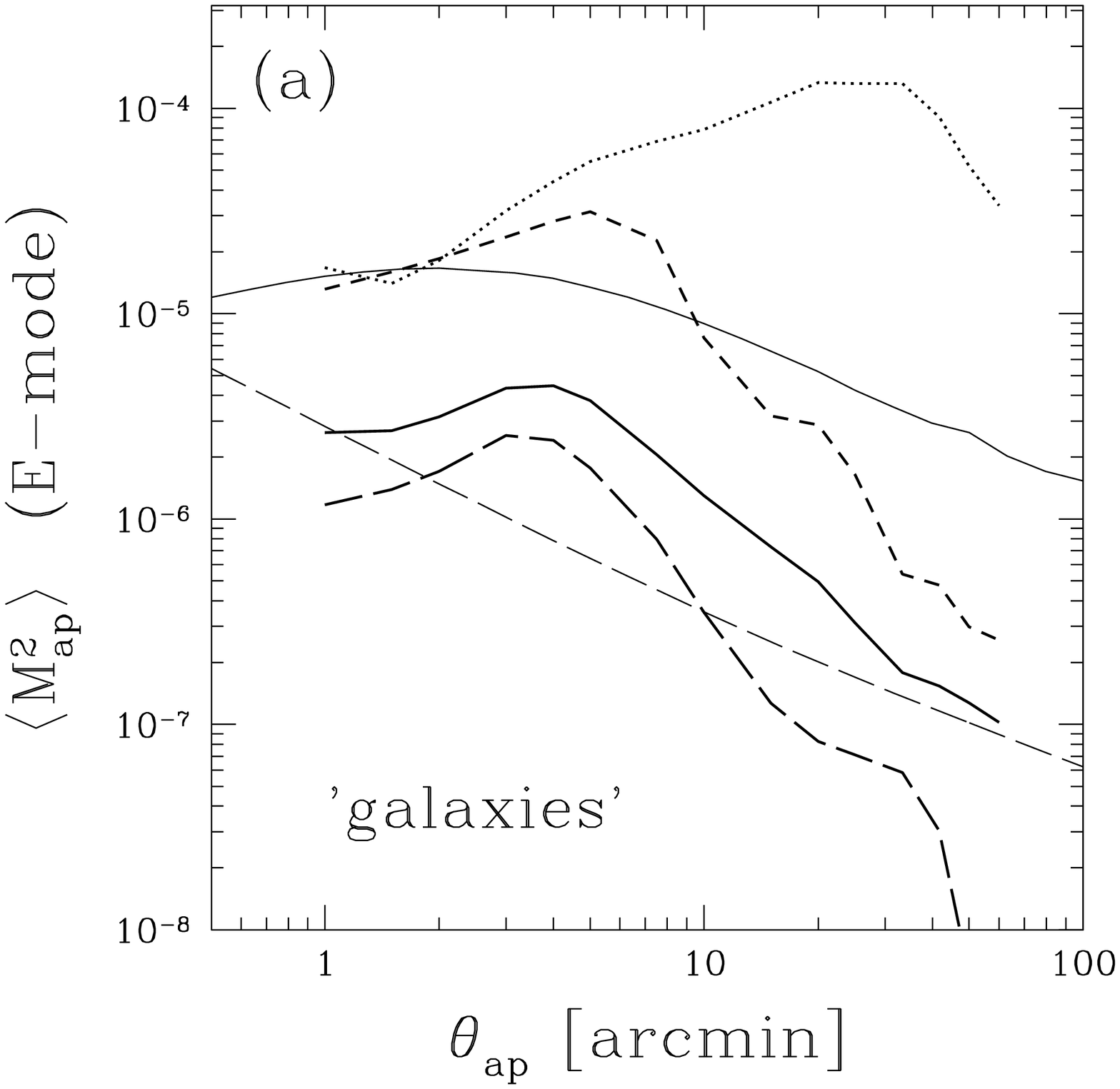}
\epsfxsize=0.48\hsize
\epsfbox{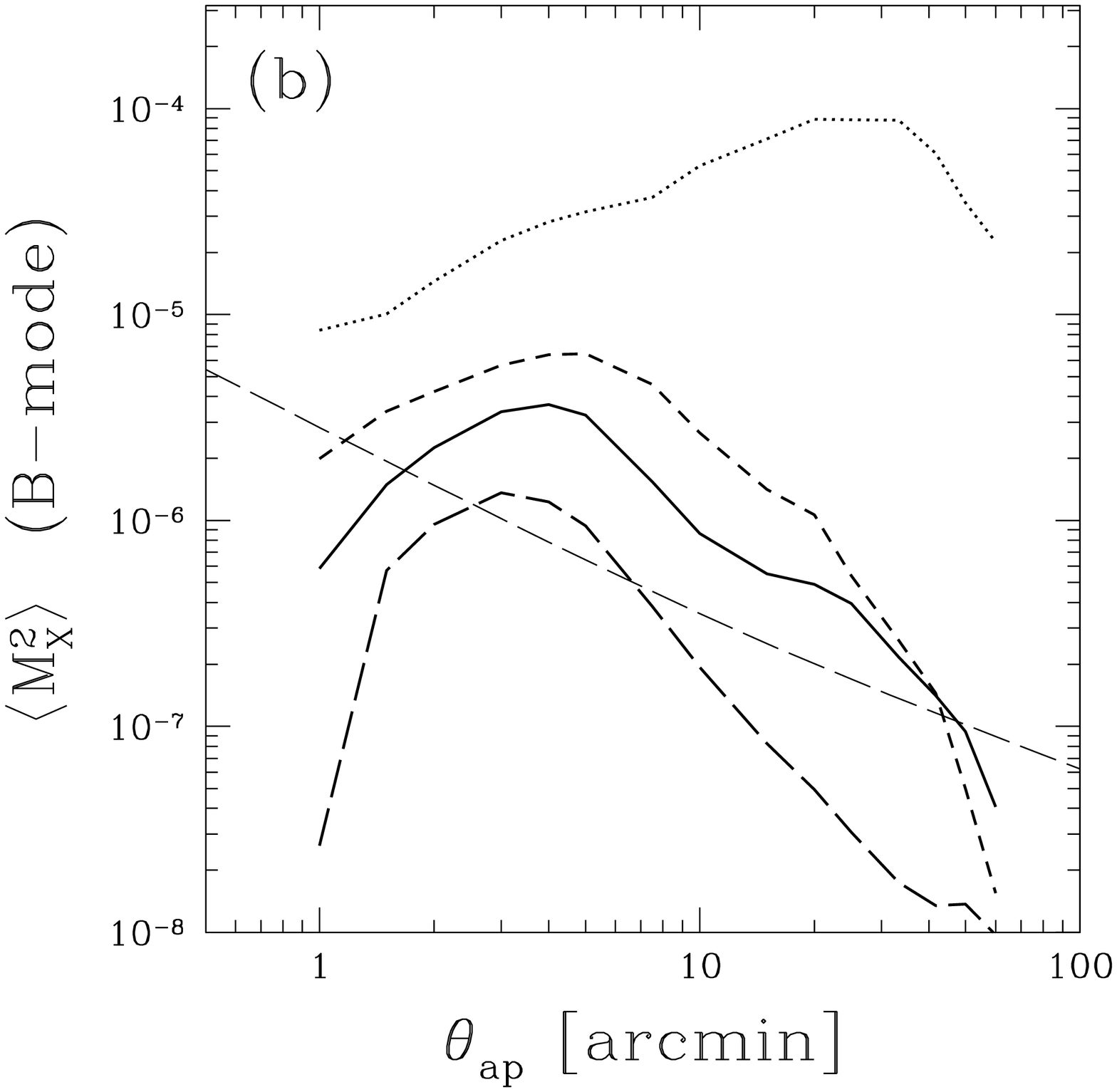}}
\end{center}
\vspace{-0.5cm}
\caption{\small (a) The expected amplitude of the variance in the aperture
mass statistic $M_{\rm ap}$ (``E''-mode) as a function of aperture
size in deep observations of galaxies. (b) The expected amplitude of
the ``B''-mode as a function of aperture size.  The dotted curve
indicates the signal without any correction for PSF anisotropy. The
short dashed line corresponds to the results when a second order polynomial
is used to characterize the PSF anisotropy (standard approach).  If a
field with a large density of stars is observed to obtain an accurate
model for the PSF anisotropy the residual signal can be significantly
reduced as is indicated by the solid line. Even better results are
obtained by deriving a detailed model from a field with many stars,
and using a scaled version of this model with additional low order
components, as is indicated by the long dashed lines. The thin solid line
indicates the expected amplitude of the cosmic shear signal from the
CFHTLS and the thin dashed line indicates the projected $1\sigma$
statistical error. The measurements will be dominated by systematics
if the standard correction approach is used, but much better results
are expected with an improvement in the modelling of the spatial
variation of the PSF anisotropy.
\label{psf_an}}
\end{figure}

\subsection{Dealing with systematics}

The distortion in the images caused by the intervening large scale
structure is typically less than a percent in amplitude, much less
than the intrinsic shapes of the galaxies themselves. Hence, the
observed shape of each galaxy provides only a very noisy estimate of
the lensing signal. This is the reason why the commissioning of
panoramic cameras has played a major role in this field: large numbers
of galaxies are required for a good measurement of the lensing signal.
In the absence of systematic distortions, the galaxy shape provides an
unbiased estimate of the lensing signal. Unfortunately, in practice,
various observational effects change the shapes of the galaxies.

The seeing circularizes the images, thus reducing the amplitude of the
lensing signal. To minimize the effect of seeing, lensing surveys
require good imaging conditions. Furthermore, the PSF is never
perfectly round, and the PSF anisotropy causes coherent alignments in
the galaxy shapes, in principle mimicking a lensing signal. Finally
the camera optics typically distort the images. The latter is usually
well controlled, and can be corrected for using a good astrometric
solution for the observed field.

The correction for the PSF has been studied in great detail and in
recent years several new approaches to deal with the PSF have been
proposed (e.g., Kuijken 1999; Kaiser 2000; Bernstein \& Jarvis 2002;
Refregier \& Bacon 2003). An early method was developed by Kaiser et
al. (1995), which approximates the PSF as the convolution of a
axisymmetric component with a compact, anisotropic kernel. This
technique separates the correction for PSF anisotropy and the
correction for the circularisation of the PSF.  The latter is
discussed in Luppino \& Kaiser (1997). Although the simplifying
assumptions of Kaiser et al. (1995) are typically not valid for real
data, the method has proven to work remarkably well for the first
generations of cosmic shear surveys. It is not clear, however, whether
the Kaiser et al. (1995) approach is accurate enough for the upcoming
surveys.

To correct for the PSF, one selects a sample of moderately bright
stars and measures their shape parameters. A model is fitted to these
measurements to characterize the spatial variation of the PSF
anisotropy. This model is then used to correct the shapes of the
galaxies. Typically a second order polynomial is used to describe the
variation in the PSF anisotropy, because higher order polynomials are
unstable due to the finite number of stars in the data. 

Recently Hoekstra (2004) examined whether the adopted model for the
variation of the PSF anisotropy can result in systematic
errors. Hoekstra (2004) argues that one expects small errors on large
scales, because the residual signal will occur on scales smaller than
the size of the chips on the camera. The small scale measurements,
however, can be seriously affected. Note that this is the case even if
the scheme to correct for the PSF anisotropy is perfect! Hoekstra
(2004) considered data taken with the CFH12k camera and found that a
second order polynomial gives rise to a residual signal, with an
``E''-mode which is $\sim 3$ times the ``B''-mode. Different
parameterisations, however, can significantly reduce residual
systematics, as is shown in Figure~\ref{psf_an}. This finding is
particularly relevant for the VIRMOS-DESCART survey (van Waerbeke et
al. 2001,2002) which shows a similar anisotropy pattern.

\section{Observational results}

In this section we review some of the recent results from cosmic shear
surveys. We note, however, that the list of results discussed here is
incomplete. The field has evolved so quickly, that it is not very
useful to discuss some of the early results. For instance, the way
residual systematics are quantified, using the ``E'' and ``B''-mode
decomposition, has only recently been implemented.

\begin{table}[!b]
\begin{center}
\begin{tabular}{lcccl}
\hline
\hline
paper & $\sigma_8$ & area & $m_{\rm lim}$ & telescope \\
      &            & [deg$^2$]    & [mag]     &  \\
\hline
van Waerbeke et al. (2001) & $0.88\pm0.11$ & 8 & I=24 & CFHT \\
Rhodes et al. (2001) & $0.91^{+0.25}_{-0.29}$ & 0.05 & I=26 & HST \\ 
Bacon et al. (2002) & $0.97\pm0.13$ & 1.6 & R=25 & Keck/WHT \\
Refregier et al. (2002) & $0.94\pm0.17$ & 0.36 & I=23.5 & HST \\
van Waerbeke et al. (2002) & $0.94\pm0.12$ & 12 & I=24 & CFHT \\
Hoekstra et al. (2002b) & $0.86^{+0.04}_{-0.05}$ & 53 & R=24 & CFHT/CTIO\\
Brown et al. (2003) & $0.74\pm0.09$ & 1.25 & R=25.5 & ESO \\
Hamana et al. (2003) & $0.69^{+0.35}_{-0.25}(2\sigma)$ & 2.1 & R=26 & Subaru\\
Jarvis et al. (2003) & $0.71^{+0.12}_{-0.16}(2\sigma)$ & 75 & R=23 & CTIO\\
\hline
\hline
\end{tabular}
\begin{footnotesize}
\caption{Summary of recent cosmic shear results. The various surveys
show a large range in survey area, telescopes, filters and
depth. Therefore it is difficult to compare the actual lensing
signals. However, one compare the results by considering the value of
$\sigma_8$ (the normalisation of the power spectrum) for a fiducial
value of $\Omega_m=0.3$. Although there are some issues concerning the
adopted source redshift distributions and quoted error bars (partly
because some studies do not marginalize over other cosmological
parameters), the overall agreement between the various measurements is
quite good. The more recent results typically use the decomposition into
``E'' and ``B''-modes, allowing them to estimate the level of systematics.
Note that the value from the VIRMOS-DESCART survey 
(van Waerbeke et al. 2002) will decrease somewhat using the revised cosmic 
shear signal.\label{tab_s8}}
\end{footnotesize}
\end{center}
\end{table}

Table~\ref{tab_s8} lists a summary of recent cosmic shear results. The
table shows that the various surveys span a large range in area,
depth.  In addition, different filters and telescopes have been used.
Consequently, one cannot simply compare the amplitudes of the lensing
signals. Instead it is common to compare the normalization of the
matter power spectrum $\sigma_8$ at a fiducial value of $\Omega_m=0.3$
(using a flat geometry). In principle, this comparison removes the
dependence of the lensing signal on the source redshift
distribution. However, various groups use different distributions, or
account differently for the uncertainty in the redshift
distributions. Furthermore, the determination of $\sigma_8$ also
depends slightly on other parameters, such as the shape parameter
$\Gamma$ and the predicted non-linear power spectrum. Despite all
these differences, the overall agreement in the derived values for
$\sigma_8$, listed in Table~\ref{tab_s8} is quite good.

We also note that the reanalysis of the VIRMOS-DESCART survey lowers
the value of $\sigma_8$ compared to the value listed in van Waerbeke
et al. (2002). The new results show no significant ``B''-mode, and are
in excellent agreement with the results from Hoekstra et al. (2002b).
The new VIRMOS-DESCART measurements demonstrate that the current
correction techniques are adequate for surveys of a few tens of
degrees.

\begin{figure}[!b]
\begin{center}
\epsfxsize=0.9\hsize
\epsfbox[0 140 600 500]{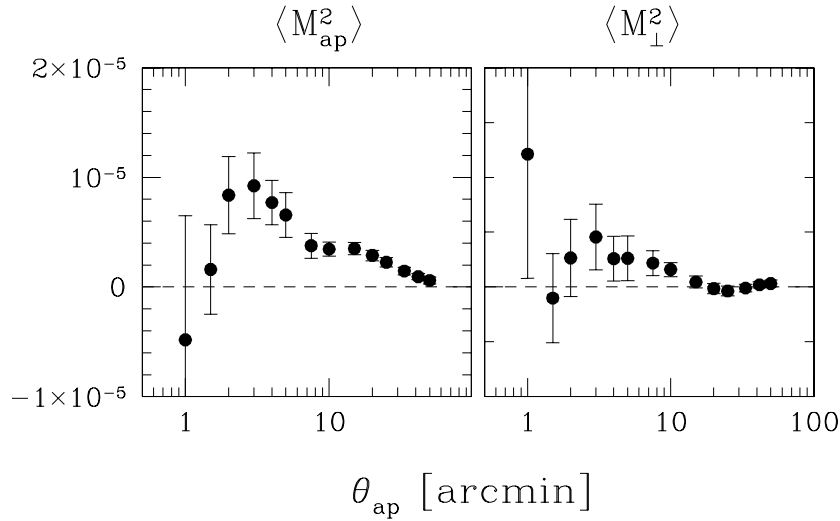}
\end{center}
\vspace{-0.5cm}
\caption{\small The left panel shows the measured variance of the aperture
mass $\langle M_{ap}^2\rangle$ as a function of aperture size using source
galaxies with $22<R_C<24$. This signal corresponds to the ``E''-mode. The
corresponding ``B''-mode or $\langle M_{\perp}^2\rangle$ is presented in 
the right panel. Note that the points are slightly correlated. The RCS
data show a significant ``B''-mode on scale $5-10$ arcminutes, which
might be caused by the intrinsic alignments of the sources. On larger
scales the ``B''-mode vanishes, and the latter measurements carry most
weight in the determination of the cosmological parameters.
\label{map_rcs}}
\end{figure}

The good agreement between the RCS measurements, shown in
Figure~\ref{map_rcs}, and the VIRMOS-DESCART data support the
cosmological origin of the signal: the amplitude scales as expected
with source redshift. Also the scale dependence is in good agreement
with what is expected from CDM simulations. The RCS data, however, do
show a (small) B-mode on small scales (right panel of
Figure~\ref{map_rcs}), which might be caused by intrinsic alignments
of the sources, which is expected for a survey this shallow. The large
scale measurements, however, appear free of systematics, and these
points carry most weight in the cosmological parameter estimation.

As discussed above, in order to relate the measurements presented in
Figure~\ref{map_rcs} to estimates of cosmological parameters, we need
to assume a source redshift distribution. For this we have used
photometric redshift distributions from the HDF North and South (see
Hoekstra et al. 2002a, 2002b for details). 

So far, most weak lensing studies have focussed on the two-point
statistics, for which the relevant parameters are the effective shape
parameter $\Gamma_{\rm eff}$ (which includes the spectral index), the
matter density $\Omega_m$ and the normalisation of the power spectrum
$\sigma_8$. Current surveys cannot place strong constraints on
$\Gamma_{\rm eff}$, and there is a strong degeneracy between
$\sigma_8$ and $\Omega_m$. Larger surveys, which probe larger scales,
and provide better measurements in the non-linear regime, will break
these degeneracies. Similar degeneracies, however, exist for the
current studies using cluster abundances.

\begin{figure}[!t]
\begin{center}
\hbox{
\epsfxsize=0.48\hsize
\epsfbox{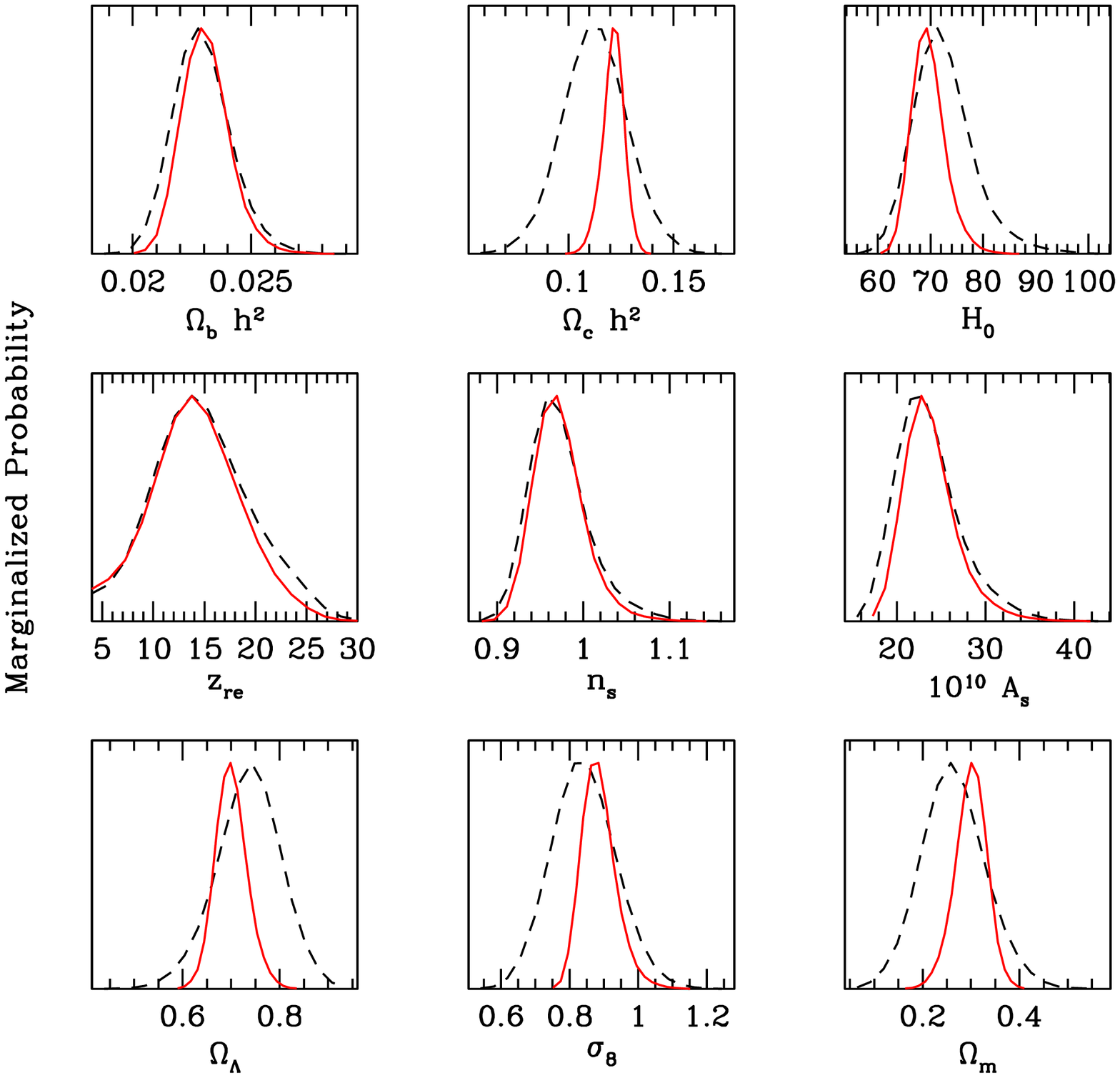}
\epsfxsize=0.48\hsize
\epsfbox{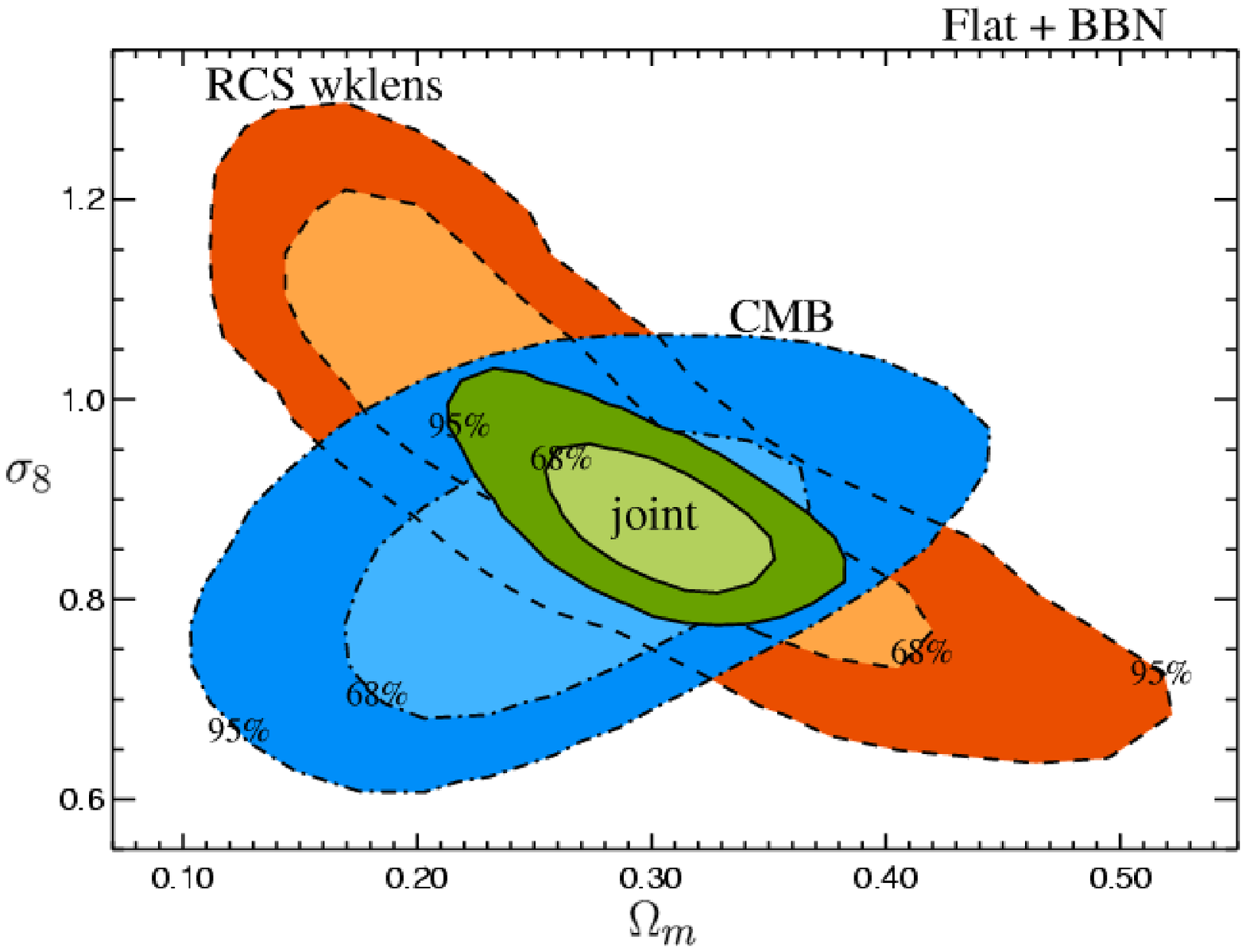}}
\end{center}
\caption{\small {\it left panel}: one dimensional marginalized
probability distribution for a selection of parameters, assuming a
flat geometry of the universe from Contaldi et al. (2003). The dashed
curves correspond to the CMB only results, whereas the solid curves
are for the combination of CMB and RCS measurements. This demonstrates
the excellent agreement between the CMB and lensing measurements, and
allows us to improve the constraints on some parameters considerably.
{\it right panel}: Joint likelihoods for $\Omega_m$ and
$\sigma_8$. This figure clearly shows the usefulness of combining CMB
and weak lensing measurements, as they are (almost)
orthogonal.\label{wmap}}
\end{figure}

Degeneracies between parameters also exist for CMB measurements, such
as the results obtained from WMAP (Bennet et al. 2003; Spergel et
al. 2003; Verde et al. 2003). The uncertainties in some parameters can
be reduced significantly by combining the CMB results with
measurements of the matter power spectrum. To this end, Verde et al.
(2003) used measurements from the 2dF Field Galaxy Redshift Survey
(e.g, Colless et al. 2001) and the power spectrum inferred from the
Ly-$\alpha$ forest (Croft et al. 2002). The uncertainty in the bias
parameter, however, limits the accuracy of the latter approach.

The weak lensing measurements can be of great use, as was shown in
Contaldi et al. (2003): the combination of current cosmic shear
surveys with WMAP results provides constraints on cosmological
parameters that are comparable (or even better) than the largest
redshift surveys. Some of the results from Contaldi et al. (2003) are
shown in Figure~\ref{wmap}. There is excellent agreement between the
lensing measurements and the CMB data. Furthermore, the constraints in
the $\Omega_m-\sigma_8$ plane are (almost) orthogonal.

The constraints from redshift surveys are not expected to improve
dramatically in the coming years (the constraints from SDSS will
tighten constraints compared to 2dF). Constraints from cosmic shear
surveys, on the other hand, will improve dramatically. The
signal-to-noise ratio of the improved VIRMOS-DESCART measurements is
already much better than the RCS results shown here. In the next five
years, the CFHT Legacy Survey (see \S5) will survey an area 10 times
that of the VIRMOS-DESCART survey. These developments, in conjunction
with the improvement in CMB measurements, as well as other probes,
make cosmic shear an important tool in ``precision cosmology''.

\section{Galaxy biasing}

An alternative way to derive constraints on the matter power spectrum
is to use the distribution of observable galaxies to infer the
underlying matter distribution, under the assumption that light traces
matter. On large scales the latter is likely to be true, but the
drawback of galaxy redshift surveys is that the galaxies are biased
tracers: the amplitude of the galaxy power spectrum is $b^2$ times the
matter power spectrum (i.e., linear biasing), where $b$ is the bias
parameter. On small scales (less than a few Mpc) the situation becomes
even more unclear.  Galaxy formation is a complex process, with many
(unknown) feedback processes regulating the formation of stars. On
these scales the galaxy bias can be non-linear, scale dependent or
stochastic. This poses a problem for the interpretation of the results
from galaxy redshift surveys, but can provide important observational
clues for our understanding of galaxy formation.

Previously, observational constraints on the bias parameter $b$ came
from measurements of the galaxy two-point correlation function, which
is compared to the matter correlation function computed from numerical
simulations. Such studies suggest that $b$ varies with scale, but the
results rely on the assumptions made for the numerical simulations.
Furthermore, this procedure cannot be used to examine how tight the
correlation between the light and the matter is. To do so, we need to
measure the galaxy-mass cross-correlation coefficient $r$, which is a
measure of the amount of stochastic and non-linear biasing (e.g., Pen
1998; Dekel \& Lahav 1999).

Weak lensing provides the most direct way to measure the galaxy-mass
correlation function (e.g., Fischer et al. 2000; McKay et al. 2001;
Hoekstra et al. 2003).  To study the galaxy biasing we use a
combination of the galaxy and mass auto-correlation functions, as well
as the cross-correlation function. The bias parameters are defined in
terms of the observed correlation functions through (Hoekstra et
al. 2002c)

\begin{equation}
b^2=f_1(\theta_{\rm ap},\Omega_m,\Omega_\Lambda)\times
\frac{\langle{N}^2(\theta_{\rm ap})\rangle}
{\langle M_{\rm ap}^2(\theta_{\rm ap})\rangle},
\end{equation}

\noindent and

\begin{equation}
r=f_2(\theta_{\rm ap},\Omega_m,\Omega_\Lambda)\times
\frac{\langle M_{\rm ap}(\theta_{\rm ap}){N}(\theta_{\rm ap})\rangle}
{\sqrt{\langle {N}^2(\theta_{\rm ap})\rangle
\langle M_{\rm ap}^2(\theta_{\rm ap})\rangle}},
\end{equation}

\noindent where $M_{\rm ap}(\theta_{\rm ap})$ is the aperture mass
obtained from lensing (see Eqn.~2), and $N(\theta_{\rm ap})$
corresponds to the aperture galaxy counts (filtered with the
compensated filter; see Hoekstra et al. 2002c for details). The
functions $f_1$ and $f_2$ depend on the assumed cosmological model and
the redshift distributions of the lenses and the sources (for details
see Hoekstra et al. 2002c). The values of $f_1$ and $f_2$, however,
depend minimally on the assumed power spectrum and the angular
scale. There is no reason for $b$ or $r$ to be constant with scale,
but as has been shown in Hoekstra et al. (2002c) we can actually
measure the bias parameters as a function of scale using Eqs.~6 and~7.

\begin{figure}[!t]
\begin{center}
\epsfxsize=0.7\hsize
\epsfbox{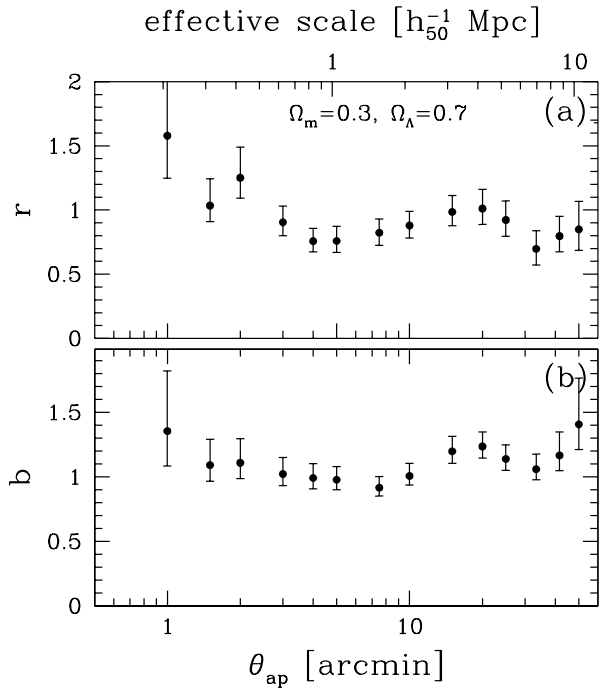}
\end{center}
\vspace{-0.5cm}
\caption{\small (a) The measured value of the galaxy-mass cross-correlation
coefficient $r$ as a function of the scale for a $\Lambda$CDM 
$(\Omega_m=0.3; \Omega_\Lambda=0.7)$ cosmology. (b) The bias parameter
$b$ as a function fo scale. The upper axis indicates the effective
physical scale probed by the compensated filter at the median
redshift of the lenses $(z=0.35)$. The error bars correspond
to the 68\% confidence intervals. Note that the measurements at
different scales are slightly correlated.
\label{bias}}
\end{figure}

Hoekstra et al. (2002c) studied the bias properties of galaxies with
$19.5<R_C<21$. They combined the weak lensing measurements from the
RCS and the VIRMOS-DESCART survey (e.g., van Waerbeke et al 2002) to
obtain the first direct measurements of the bias parameter $b$ and the
galaxy-mass cross-correlation coefficient $r$ as a function of scale
(with scales ranging from 0.1 to 4 $h^{-1}$ Mpc).  As discussed above,
the VIRMOS-DESCART signal suffered from residual systematics, and 
Figure~\ref{bias} shows the bias parameters as a function of scale,
using the improved cosmic shear signal. The reduced $B$-mode on large
scales significantly reduced the uncertainties in the bias parameters.
In addition the variation with scale (around $\sim 5$ arcmin.) is 
less pronounced.

Multi-color data for the RCS have been obtained and the first catalogs
have been created. The first results suggest that reliable photometric
redshifts can be derived. Consequently one can start to examine the
bias parameters as a function of color and luminosity, which also
simplifies the comparison with predictions from models of galaxy
formation.

\section{Prospects}

Since the first detections of lensing by large scale structure, only a
few years ago, this field in observational cosmology has advanced
significantly: detections have become measurements! The combination of
cosmic shear measurements with CMB data already give some of the best
constraints on cosmological parameters. It appears that, at least for
now, most of the systematic errors are under control.

Despite these exciting recent developments, it is clear that the next
generation of cosmic shear surveys requires improvements in various
areas in order to be a successful tool in this era of ``precision
cosmology''. The Canada-France-Hawaii-Telescope Legacy Survey (CFHTLS)
covers an area 10 times that of the VIRMOS-DESCART data set. Somewhat
smaller surveys will be done by other telescopes (e.g., Deep Lens
Survey, Subaru). This will result in a significant reduction in the
statistical uncertainties.  However, it is not clear that the current
techniques to correct for the PSF are sufficiently accurate, but it
seems plausible that the correction for the PSF can be dealt
with. More problematic is the issue of the source redshift
distribution. To date, redshift surveys have targeted only small areas
of the sky. Some new surveys aim to cover larger areas, but the
timescales for those are probably larger than the progress expected in
weak lensing. Fortunately, the new lensing surveys will have
photometric redshift information, which will reduce the uncertainty in
the redshift distribution.

Given the above, it seems reasonable to assume that systematics can be
reduced to the required level on the observational side. On the
theoretical side improvements are required as well: much of the
sensitivity to cosmological parameters comes from the measurements on
the non-linear power spectrum. However, the current accuracy with
which the latter is known is comparable to the statistical
uncertainties of the VIRMOS-DESCART survey. A proper interpretation of
the CFHTLS measurements will require a prediction for the non-linear
power spectrum with an accuracy $\sim 1\%$ or better.

Although the analysis of the next generation of cosmic shear surveys
will be a challenging task, much more ambitious projects have been
proposed already. For instance the Large aperture Synoptic Survey
Telescope (LSST; Tyson et al. 2002) or Panoramic Survey Telescope and
Rapid Response System (Pan-STARRS; Kaiser et al. 2000) will produce
massive data sets from the ground. Space based observations minimize
the obervational systematics, and a weak lensing survey using the
proposed SuperNova Acceleration Probe (SNAP) can provide some of the
most accurate constraints on the equation of state of the universe.

\noindent To summarize these proceedings in a single sentence:
{\it the prospects for weak lensing are excellent!}

\end{document}